\documentclass[%
 reprint,
 amsmath,amssymb,
 aps,
]{revtex4-2}

\usepackage{graphicx}
\usepackage{dcolumn}
\usepackage{bm}
\usepackage{float}

\begin{document}

\preprint{APS/123-QED}

\title{Effect of RKKY and dipolar interaction on the nucleation of skyrmion in Pt/Co multilayer with Ir spacer}
\author{Shaktiranjan Mohanty$^{1}$, Brindaban Ojha$^{1}$, Bhuvneshwari Sharma$^{1}$, Ashutosh Rath$^{2}$, Chandrasekhar Murapaka$^{3}$ and Subhankar Bedanta$^{1,4, *}$}

\affiliation{$^{1}$Laboratory for Nanomagnetism and Magnetic Materials (LNMM), School of Physical Sciences, National Institute of Science Education and Research (NISER), An OCC of Homi Bhabha National Institute (HBNI), Jatni, Odisha 752050, India.}
\affiliation{$^{2}$CSIR-Institute of Minerals and Materials Technology (IMMT), Bhubaneswar, 751013, India}
\affiliation{$^{3}$Department of Materials Science and Metallurgical Engineering, Indian Institute of Technology-Hyderabad, Kandi, Telangana 502284, India}
\affiliation{$^{4}$Center for Interdisciplinary Sciences(CIS), National Institute of Science Education and Research (NISER),  An OCC of Homi Bhabha National Institute (HBNI), Jatni 752050, Odisha, India}

\email{sbedanta@niser.ac.in}


\begin{abstract}
Magnetic skyrmions, topologically protected spin textures, have emerged as promising candidates for next-generation spintronic applications. In this study, we investigate the stabilization of skyrmionic states in a uniquely engineered Pt/Co multilayer system with an Ir spacer, where both Ruderman–Kittel–Kasuya–Yosida (RKKY) and dipolar interactions play a crucial role. The studied multilayer structure consists of a synthetic antiferromagnetic (SAF) configuration, where a single Ir layer facilitates strong antiferromagnetic coupling between two ferromagnetic regions: FM1 (top) and FM2 (bottom), each formed by repeated Co layers separated by Pt, enabling significant dipolar interactions. This FM1/Ir/FM2 configuration results in a distinctive skyrmionic hysteresis loop, driven by the interplay of dipolar and RKKY interactions. Magnetic force microscopy (MFM) imaging confirms the nucleation of isolated skyrmions, while magnetotransport measurements reveal a finite topological Hall effect (THE), indicating the chiral nature of these spin textures. Furthermore, we demonstrate that increasing the Co layer thickness leads to a reduction in magnetic anisotropy, which in turn results in the formation of relatively larger and denser skyrmions. Our findings establish a robust approach for stabilizing skyrmions through the combined effects of dipolar and RKKY interactions, offering new pathways for controlled skyrmion manipulation in spintronic devices.
\end{abstract}

\maketitle


\section{\label{sec:level1}Introduction}

The emergence of magnetic skyrmions as topologically protected spin textures has sparked significant interest in the field of condensed matter physics and spintronics due to their unique properties and potential applications in future magnetic storage and computing devices \cite{fert2017magnetic, wang2022fundamental, jonietz2010spin, schulz2012emergent, iwasaki2013universal}. These nanoscale spin configurations, characterized by a swirling arrangement of magnetic moments, exhibit topological stability and non-trivial spin textures, making them promising candidates for information storage and manipulation at ultrahigh densities and low energy consumption  \cite{fert2013skyrmions, nagaosa2013topological, sampaio2013nucleation, iwasaki2013current}. In recent years, considerable efforts have been devoted to the investigation of skyrmion nucleation and manipulation in various magnetic systems, including thin films, multilayers, and heterostructures  \cite{soumyanarayanan2017tunable, he2017realization, lin2018observation, cheng2023room, ojha2023driving, boulle2016room}. Characterising the spin configuration of these skyrmions is one of the primary objectives to comprehend the underlying competing interactions that stabilise them as well as for optimizing device performance and designing novel skyrmion-based technologies. Previously, observation of skyrmions has been extensively studied in thin film multilayer systems with the repetitions of heavy metal (HM)/ ferromagnet (FM) layers  \cite{woo2016observation, moreau2016additive, salikhov2022control, lemesh2018twisted}, by considering the FM thickness near to spin reorientation transition (SRT)  \cite{herve2018stabilizing, jiang2015blowing}, with enhanced DMI by taking different heavy metal (HM) at both interfaces of FM layer  \cite{tolley2018room}, etc. Among the various thin film systems hosting chiral spin textures like skyrmions, HM1/FM/HM2 magnetic multilayers have garnered significant interest due to the generation of non-cancelling interfacial DMI produced by breaking the inversion symmetry. Also, it provides the flexibility to tune the magnetic properties largely by changing the thickness of each layer. He et. al., have shown the evolution of skyrmions in a Pt/Co/Ta multilayer system with Co thickness near to SRT  \cite{he2017realization}. Lin et. al., have shown the observation of these chiral spin textures in a Pt/Co/W multilayer system with enhanced Dzyaloshinskii-Moriya interaction (DMI)  \cite{lin2018observation}. Recently, Cheng et. al., have demonstrated the observation of bubble-like Neel skyrmions in a Pt/Co/Cu multilayer system by tuning the Co thickness as well as the number of periods of these multilayers  \cite{cheng2023room}. However, there are very few reports which demonstrate the nucleation of these skyrmionic features by considering the FM thickness much below the SRT (in the PMA regime), with less repetition of number of periods of these multilayers and simultaneously introducing the Ruderman–Kittel–Kasuya–Yosida (RKKY) interaction between the FM layers. In our previous report  \cite{mohanty2024observation}, we have shown the stabilization of skyrmionic states in a Pt/Co/Ir/Co/Pt system where, the two Co layers below and above the Ir spacer layer are not antiferromagnetically (AFM) coupled at that particular thickness of Ir. In this work, we have considered a particular thickness of Ir spacer to have strong AFM coupling between the Co layers adjacent to the spacer which forms a synthetic antiferromagnetic (SAF) structure. SAFs are basically two FM layers coupled antiferromagnetically via non-magnetic (NM) spacer layer  \cite{duine2018synthetic, mohanty2022magnetization, gabor2017interlayer}. SAFs are proved to be a promising material system for hosting skyrmions which can largely reduce the skyrmion Hall effect (SkHE). The deflection of skyrmions to the edge of the track away from the applied current direction due to the topological Magnus force, leads to the SkHE  \cite{chen2017skyrmion, jiang2017direct}. By using an SAF system, the topological magnus force arising from the AFM coupled top and bottom FM layers get cancelled which reduces the SkHE  \cite{dohi2019formation, pham2024fast}. However, we have increased the number of repetitions of Pt/Co layers below and above the Ir spacer layer to simultaneously have the effect of dipolar interaction and RKKY interaction for the stabilization of skyrmions in our systems. By carefully balancing the contributions from DMI energy, RKKY interaction energy, dipolar energy, and magnetic anisotropy energy, we have achieved the nucleation of high-density skyrmions in these multilayer systems. The observation of skyrmions in these systems have been confirmed via the detection of topological Hall effect (THE) signals and magnetic force microscopy (MFM) imaging.

\section{Experimental Details}
Two samples S1 and S2 have extensively been studied in this work with structure described in the table \ref{tab1} by varying $t_{\text{Co}}$ as 0.8 and 1.0 nm and named as S1 and S2, respectively. Another two samples have also been prepared as control samples to demonstrate the existence of perpendicular magnetic anisotropy (PMA) in the Pt/Co system and AFM coupling between the Co layers via Ir spacer. The sample names with structures have been mentioned in Table 1 below. All the samples have been prepared in a high-vacuum multi-deposition chamber manufactured by Mantis Deposition Ltd., UK. The base pressure of the chamber was $\sim 7.5 \times 10^{-8}$ mbar. The deposition pressure was $\sim 1.7 \times 10^{-3}$ mbar for Ta and $\sim 1.5 \times 10^{-3}$ for Pt, Co and Ir layers. During sample preparation, the substrate table was rotated at 15 rpm to minimize the growth-induced anisotropy and also to have uniform growth of the films. The rates of deposition were 0.1 \AA/s, 0.13 \AA/s, 0.3 \AA/s and 0.1 \AA/s for Ir, Ta, Pt and Co, respectively.

\begin{table}
\caption{Sample names with their structures. All the thicknesses shown in the parentheses are in nm.}
\label{tab1}

\resizebox{\columnwidth}{!}{
\begin{tabular}{|c|c|}
\hline
Sample name & Structure \\
\hline
R1-FM & Sub/Ta(3)/Pt(3.5)/Co(0.8)/Ir(1.3)/Pt(3.5) \\
\hline
R2-SAF & Sub/Ta(3)/Pt(3.5)/Co(0.8)/Ir(1.3)/Co(0.8)/Pt(3.5) \\
\hline
S1 & Sub/Ta(3)/Pt(2.5)/[Pt(1)/Co(0.8)]$_3$/Ir(1.3)/[Co(0.8)/Pt(1)]$_2$/Pt(2.5) \\
\hline
S2 & Sub/Ta(3)/Pt(2.5)/[Pt(1)/Co(1.0)]$_3$/Ir(1.3)/[Co(1.0)/Pt(1)]$_2$/Pt(2.5) \\
\hline
\end{tabular}
}
\end{table}

The schematic of the sample structure for the samples S1 and S2 has been shown in Figure 1 (a). For the structural characterization of our samples, we have performed cross-sectional TEM imaging on the sample S1 (shown in Figure 1 (b)) in a high-resolution transmission electron microscope (HR-TEM) (JEOL F200, operating at 200 kV and equipped with a GATAN oneview CMOS camera). The magnetization reversal of the samples have been measured at room temperature using a SQUID-VSM (Superconducting Quantum Interference Device-Vibrating sample magnetometer) manufactured by Quantum Design, USA. In order to elucidate the existence of skyrmionic states in both the samples, we have further performed magnetic force microscopy (MFM) measurement with an attodry 2100 MFM system manufactured by Attocube, Germany. Further, magnetotransport measurements have been performed on both samples S1 and S2 via a physical property measurement system (PPMS) manufactured by Quantum Design, USA, to confirm the observed spin textures to be chiral in nature.

\section{Results and Discussion}
Figure 1 (b) shows the high-resolution TEM image of the sample S1. From the TEM image, the growth of these ultrathin layers is clearly visible. All the respective layers with their names have been indicated in the image itself. 

\begin{figure}[H]
	\includegraphics[width=0.4\textwidth]{"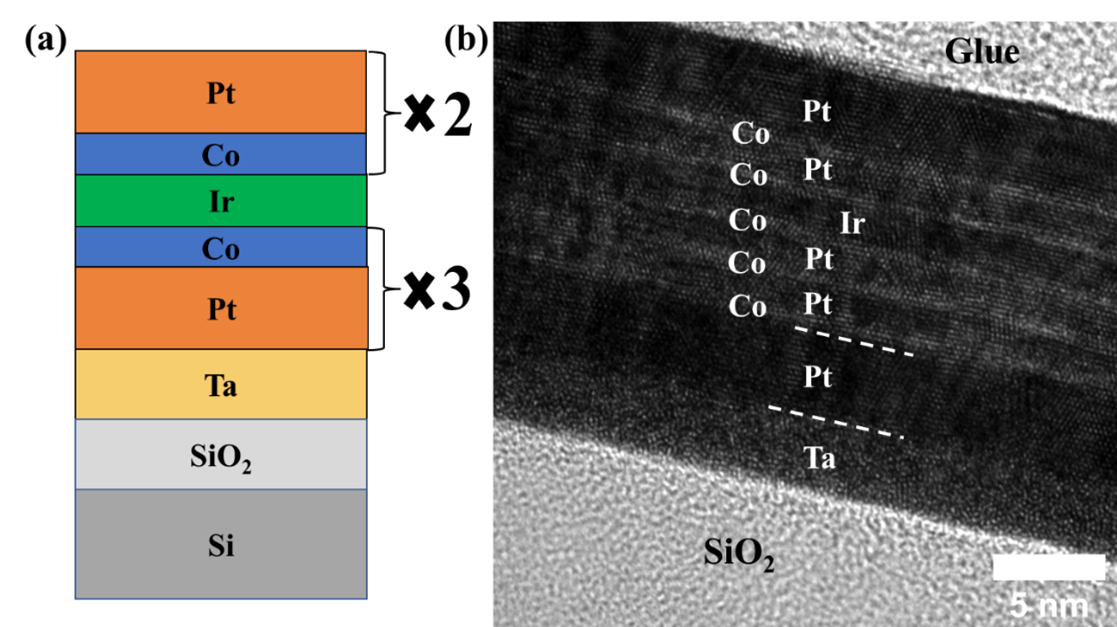"}
	\caption{(a) Schematic of the sample structure for sample S1 (tCo=0.8nm) and S2 (tCo=1.0nm), (b) Cross-sectional TEM image of the sample S1.}
	\label{fig:fig1}
\end{figure}

Magnetization reversal of the reference sample R1-FM has been shown in Figure 2 (a). Here the single Co layer with thickness 0.8nm deposited on the Pt heavy metal (HM) layer, shows strong PMA in the sample. The Ta layer has been used as a seed layer on the substrate which also favours the growth of Pt to have PMA in the Co layer. We have used the same thickness of Co i.e., 0.8nm for S1. However, for sample S2, the thickness of Co has been considered as 1.0nm to investigate the effect of reduction in the effective anisotropy on the evolution of skyrmionic spin textures. The effective magnetic anisotropy energy density $K_{eff}$, which governs the stability of perpendicular magnetization, is given by, $K_{\text{eff}} = K_v + \dfrac{K_s}{t_{\text{Co}}}$ , where $K_v$ is the volume anisotropy and $K_s$ is the surface/interface anisotropy \cite{ikeda2010perpendicular}. As $t_{Co}$ increases, the contribution from interfacial anisotropy per unit volume $(K_s/t_{Co} )$ decreases, leading to a reduction in $K_eff$, and thus promoting a tendency toward in-plane magnetization. This makes the system more favourable for the formation of non-uniform magnetic textures such as skyrmions, especially in the presence of competing interactions like DMI and dipolar coupling. Further, before preparing the multilayer samples, we have optimized Ir spacer layer thickness to have strong AFM coupling between the Co layers. Figure 2 (b) shows the step-like behaviour in the hysteresis loop of the sample R2-SAF which indicates the AFM coupling between the FM layers forming a synthetic antiferromagnetic (SAF) structure for an Ir spacer thickness ($t_{Ir}$) of 1.3nm. 

\begin{figure}[H]
	\includegraphics[width=0.48\textwidth]{"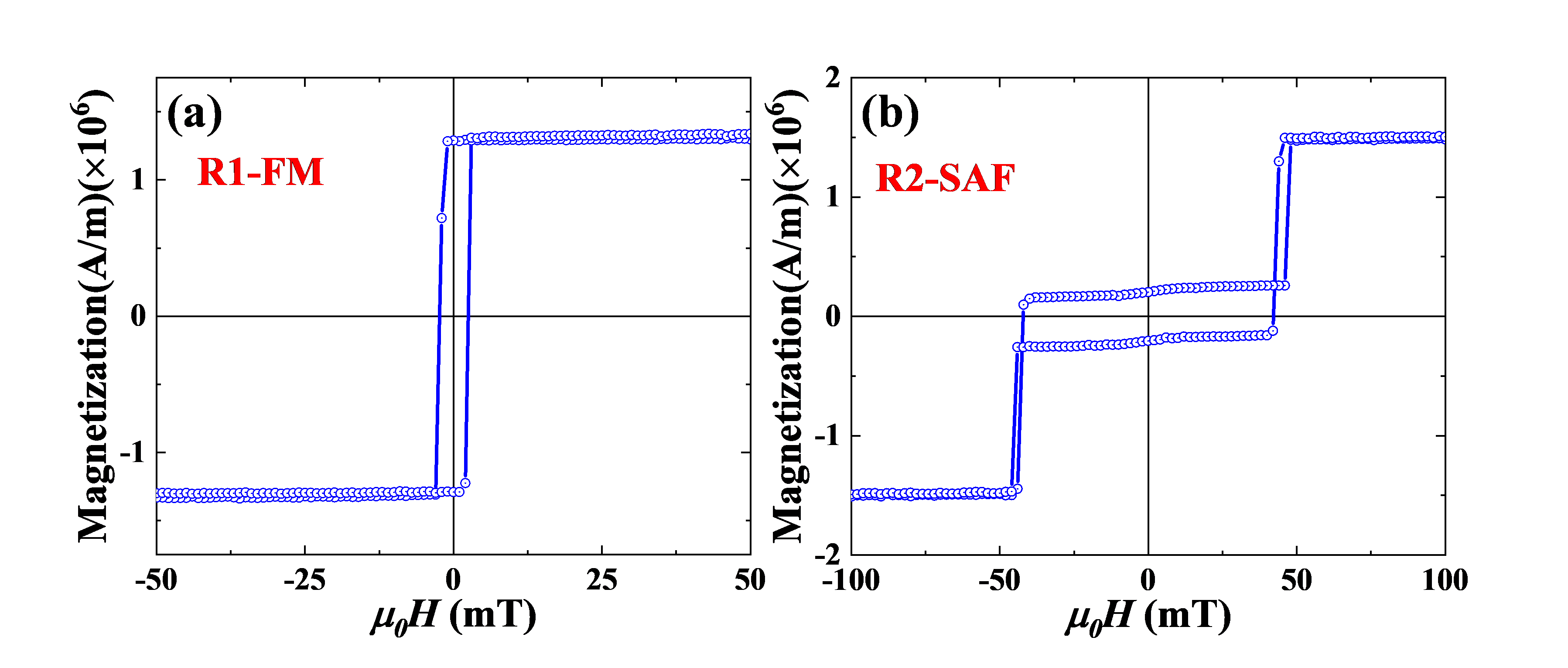"}
	\caption{Magnetization reversal at room temperature for the sample (a) R1-FM and (b) R2-SAF. The measurement has been performed in the presence of an out-of-plane applied magnetic field.}
	\label{fig:fig2}
\end{figure}

Further, we investigate the multilayer samples S1 and S2. Hysteresis loops of the samples S1 and S2 measured via SQUID magnetometer in presence of an out-of-plane applied magnetic field confirms the easy axis to be in the out-of-plane direction for both the samples (shown in the supplementary information Figure S1). Both the samples showed a slanted type of magnetization reversal with almost zero remanence. This may be due to the decrease in PMA of these sample with the introduction of RKKY coupling layer (i.e., Ir) between the Co layers. This indicates a favorable condition for the plausible presence of skyrmions in our multilayer samples. Although the $t_{Ir}$ is kept at 1.3nm to have AFM coupling among the Co layers below and above the spacer layer, no intermediate step has been observed in the hysteresis loop for both samples which indicates that the dipolar coupling due to the repetition of Pt/Co layers below and above the Ir spacer is dominant over the RKKY interaction. To get further insight into the change in anisotropy of the multilayer samples S1 and S2 as compared to the reference SAF sample, we have calculated the effective anisotropy energies ($K_{eff}$) of the samples R2-SAF, S1 and S2 by measuring the hard axis hysteresis loop in presence of an in-plane magnetic field. These hard axis loops have been shown in the supplementary information Figure S2 where the green arrow in each hysteresis plot indicates the anisotropy field ($H_K$) of the samples. The values of $K_{eff}$ are calculated by using the relation $K_{\text{eff}} = \dfrac{H_K M_S}{2}$, with $M_S$ being the saturation magnetization. The calculated values of $H_K$, $M_S$ and $K_{eff}$ for these three samples R2-SAF, S1 and S2 have been mentioned in the Table S1 of the supplementary information. It should be noted that the anisotropy energy is higher in the reference SAF sample (R2-SAF) as compared to the samples S1 and S2 having repetition of Pt/Co layers. This is due to the dominance of dipolar interaction over the PMA in the multilayer systems. However, there is also a reduction in the $K_{eff}$ value of S2 than S1. Although the sample structure as well as the number of repetitions of Pt/Co layers in both the sample are same, the reduction in the $K_{eff}$ value is due to the higher thickness of Co layer in S2 as compared to S1. 
MFM images shown in Figure 3 (a)-(e) and (f)-(j) show the gradual evolution of skyrmionics states from the demagnetized state for S1 and S2, respectively with the increase in magnetic field values. The appearance of labyrinth domains at the demagnetized state shown in Figure 3 (a), is a consequence of the balance among various competing energies present in our system such as, dipolar interaction, DM interaction, RKKY interaction, etc. Here, it should be noted that, the dominance of dipolar energy in our sample comes from the repetition of Pt/Co layers below and above the Ir spacer layer. The contribution from DMI energy comes from the Pt/Co interface as well as Co/Ir interface. Due to a smaller number of repetitions of Pt/Co layers, the dipolar energy may not have been sufficient to dominate over the exchange and anisotropy energy of the system. However, the simultaneous contribution of dipolar and RKKY interaction energy forms a labyrinth type of domains or a spin spiral state in the demagnetized state.  By applying out-of-plane magnetic field to the sample, these highly dense labyrinth type domains form into less dense stripes (shown in Figures 3 (b) and (g) for S1 and S2, respectively). Here, the Zeeman energy tries to make the spins align in the direction of applied magnetic field from the spin spiral state and forms the stripe phase. When the magnetic field is increased further, these magnetic stripes gradually break into isolated skyrmion like magnetic domains and forms a disordered skyrmion state. The formation of these distinct skyrmions have been shown in Figures 3 (d) and (i) for S1 and S2, respectively. By further increasing the magnetic field towards the saturation, the magnetic contrast of these isolated skyrmions gradually fades away (shown in supplementary information) and go to a single domain state by showing uniform contrast throughout the image. A few more MFM images have been taken at some other field values that show the detailed evolution of these skyrmion spin textures in both the sample S1 and S2, have been shown in the Figures S3 and S4 of supplementary information. In both the samples S1 and S2, these skyrmions have been observed to be stable in a wide range of magnetic field after their nucleation at different field values.

\begin{figure}[H]
	\includegraphics[width=0.48\textwidth]{"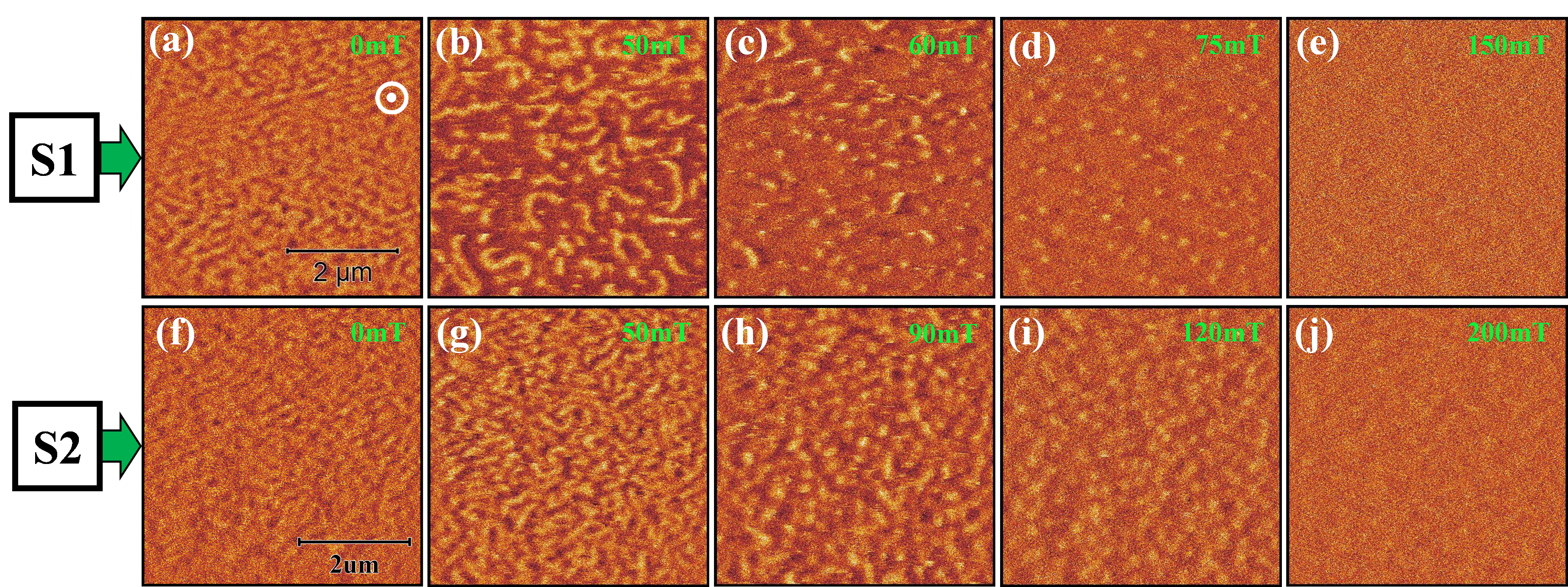"}
	\caption{MFM images of the sample S1 and S2 at different out-of-plane applied magnetic field. The field values have been mentioned in the inset of each image.}
	\label{fig:fig3}
\end{figure}

In order to confirm the observed spin textures to be chiral in nature, magneto transport measurements have been performed on both S1 and S2. Small rectangular pieces of both the samples have been measured in the Van der Pauw geometry keeping two voltage probes exactly perpendicular to the two current probes. However, we have adapted a five-probe measurement with an extra voltage probe to avoid the contribution of the additional component of the longitudinal voltage signal to the total Hall resistivity of our sample. The schematic of this five probe measurement has been shown in the Figure S5 of the supplementary information. The measurement is performed by applying a constant current of 0.5mA to the sample in the presence of a sweeping perpendicular magnetic field which results in a transverse Hall voltage proportional to the magnetic field and the charge carrier density. This gives the total resistivity of the sample which consists of these components i.e., ordinary Hall effect (OHE) and anomalous Hall effect (AHE)  \cite{nagaosa2010anomalous}. In the presence of chiral spin textures like skyrmions in the sample, an emergent magnetic field is produced due to the interaction of the conduction electrons with the skyrmions, which yields the additional contribution i.e., topological Hall effect (THE)  \cite{schulz2012emergent, neubauer2009topological, kanazawa2011large, li2013robust}. Hence, the total Hall resistivity of the system in the presence of chiral spin textures can be written as  \cite{he2018evolution, raju2019evolution, raju2021colossal, mourkas2021topological, maccariello2018electrical},
\begin{equation}
\rho_{xy} = R_0 H + R_s M + \rho_{\mathrm{THE}} \,,
\label{eq:hall}
\end{equation}
$R_0 H$ is the contribution from OHE. $R_S M$ is the contribution from AHE. The subscript $xy$ indicates the resistivity of the sample has been measured by applying the current in the x direction and measuring the voltage in the y direction. $R_0$ is the ordinary Hall coefficient, $R_s$ is the AHE coefficient, and M is the out-of-plane magnetization of the sample. By subtracting the contribution of OHE and AHE from the total Hall resistivity, the topological Hall resistivity of the sample is calculated. The curve with blue circles in Figure 4 (a) and (b) depicts the Hall resistivity (AHE and THE ($\rho_{xy}^{\mathrm{AHE+THE}}$) for our samples S1 and S2, respectively. The contribution from OHE, which is very small in our case, has been subtracted by correcting the slope of the linear part of the total Hall resistivity ($\rho_{xy}$). Further, the contribution from AHE is scaled with the $\rho_{xy}^{\mathrm{AHE+THE}}$ by evaluating the AHE coefficient (Rs) which is plotted as the red line curve $\rho_{xy}^{\mathrm{AHE}}$ in Figures 4 (a) and (b). Thus, by subtracting the $\rho_{xy}^{\mathrm{AHE}}$ from $\rho_{xy}^{\mathrm{AHE+THE}}$, we get a hump like behaviour in both the field sweeps i.e., from +ve to -ve and from -ve to +ve, which is the contribution from THE in our samples. The topological Hall resistivity has been shown in the Figures 4 (c) and (d) for sample S1 and S2, respectively.

\begin{figure}[H]
	\includegraphics[width=0.48\textwidth]{"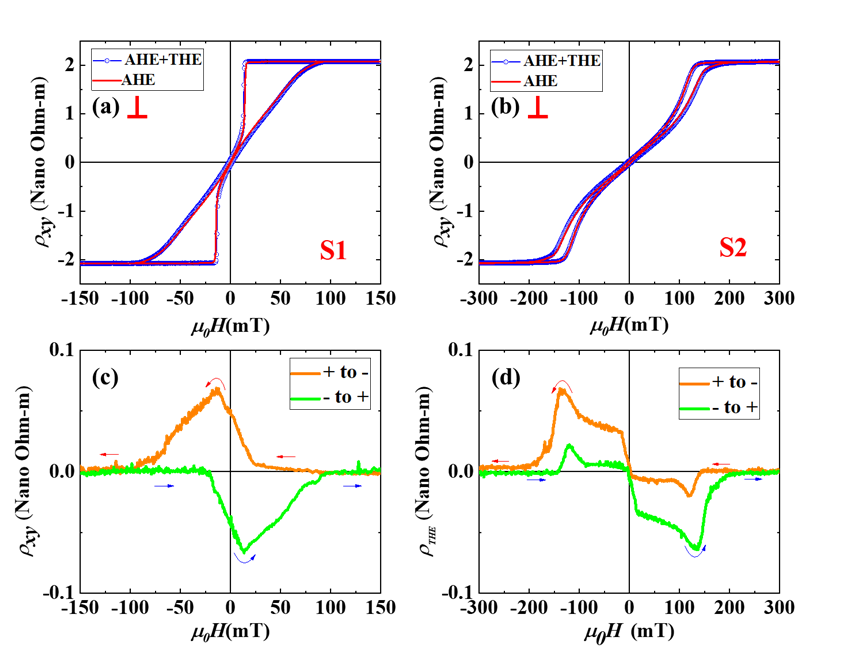"}
	\caption{Scaling of the contribution from AHE and THE (blue circle) with the AHE (red line) for (a) S1 and (b) S2. (c) and (d) represent the subtracted topological Hall resistivity for the samples S1 and S2, respectively.}
	\label{fig:fig4}
\end{figure}
The values of the topological Hall resistivity for samples S1 and S2 are found to be 0.062 and 0.068 Nano Ohm-m, respectively which is comparable to the previously reported literature values \cite{he2018evolution, raju2019evolution, mourkas2021topological}. The comparatively higher values of the THE in sample S2 as compared to S1 is due to the higher density of skyrmions observed in S2. 

\begin{figure}[H]
	\includegraphics[width=0.48\textwidth]{"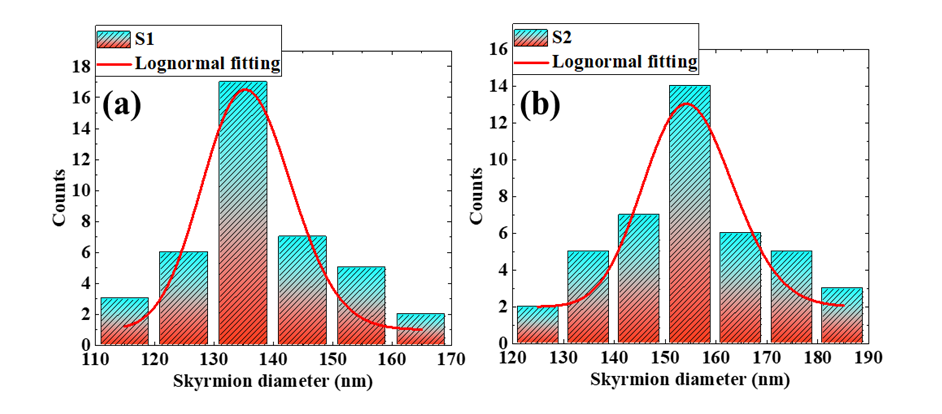"}
	\caption{The average skyrmion size analysis for the skyrmions observed in the sample (a) S1 and (b) S2. The red line in the both the plots depicts the lognormal fitting to the measured data to calculate the average skyrmion size.}
	\label{fig:fig5}
\end{figure}

By counting the number of skyrmions present in an area of $5\,\mu\mathrm{m} \times 5\,\mu\mathrm{m}$ area as shown in Figure 3, the densities of skyrmions are found to be $3.6 \times 10^{8}\ \mathrm{skyrmions/cm}^{2}$ and $5.2 \times 10^{8}\ \mathrm{skyrmions/cm}^{2}$
 for S1 and S2, respectively. Further, the shape of the skyrmions found in our sample are circular in nature. Therefore, we calculate the size of these skyrmions by measuring the diameters from the MFM images. Figure 5 (a) and (b) shows the measured diameter of various skyrmions observed in the MFM images for sample S1 and S2, respectively. From the lognormal fitting, the average skyrmion dimeter for sample S1 is found to be 135 nm and for sample S2 to be 154 nm. The increase in average size of the skyrmions in S2 is due to the reduction in the anisotropy of S2 as compared to S1  \cite{chen2020realization}.

\begin{acknowledgments}
We acknowledge the department of atomic energy (DAE), Govt. of India, for providing funding and facilities to carry out the research work. 
\end{acknowledgments}

\nocite{*}

\bibliography{References}

\clearpage
\onecolumngrid
\appendix

\section*{Supplementary Information}

\renewcommand{\thefigure}{S\arabic{figure}}
\setcounter{figure}{0}

1. Figure S1 (a) and (b) show the out-of-plane hysteresis loops for the sample S1 and S2, respectively, measured via SQUID magnetometer.

\begin{figure}[H]
\centering
\includegraphics[width=0.8\linewidth]{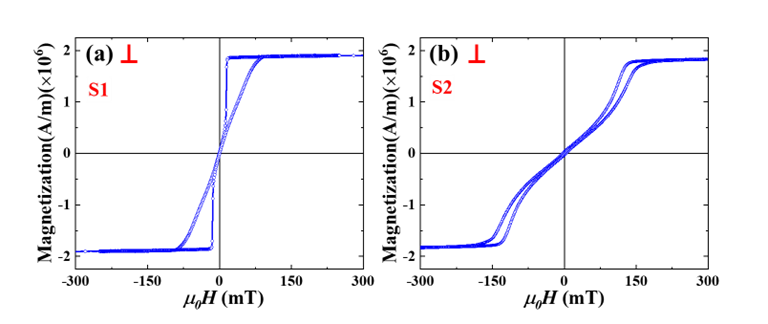}
\caption{Out-of-plane hysteresis loop for the sample (a) S1 and (b) S2, measured by SQUID-VSM.}
\end{figure}

2. The effective anisotropy energies of the samples have been calculated by measuring the hysteresis loop in presence of an in-plane applied magnetic field. Figure S2 (a), (b) and (c) show the hard axis hysteresis loops for the sample R2-SAF, S1 and S2, respectively. The green arrow in each hysteresis plot indicates the anisotropy field ($H_K$) of the samples.

\begin{figure}[H]
\centering
\includegraphics[width=0.8\linewidth]{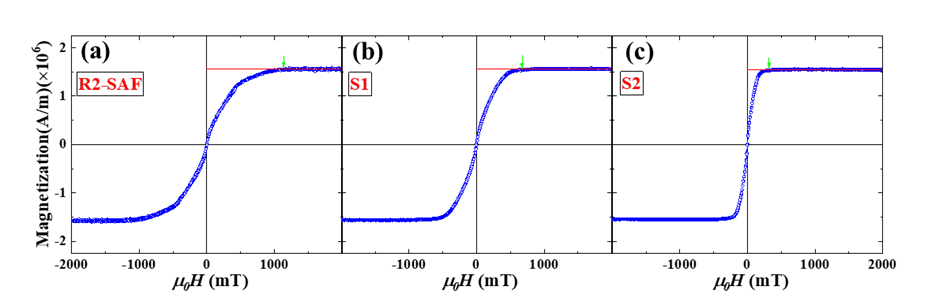}
\caption{Hard axis hysteresis loop measured in presence of an in-plane magnetic field for the samples (a) R2-SAF, (b) S1 and (c) S2. The red line in each plot starting from the y-axis and parallel to the x-axis, indicates the saturation magnetization ($M_S$) of the samples.}
\end{figure}

3. The calculated values of $H_K$, $M_S$ and $K_{\mathrm{eff}}$ for the samples R2-SAF, S1 and S2 have been mentioned in Table S1 below.

\renewcommand{\thetable}{S\arabic{table}}
\setcounter{table}{0}

\begin{table}[h!]
\caption{Calculated values of $H_K$, $M_S$ and $K_{\mathrm{eff}}$ for the samples R2-SAF, S1 and S2.}
\label{tab:S1}
\centering
\begin{tabular}{|c|c|c|c|}
\hline
\textbf{Sample} & \textbf{$M_S$ (A/m)} & \textbf{$H_K$ (mT)} & \textbf{$K_{\mathrm{eff}}$ (J/m$^3$)} \\
\hline
R2-SAF & $1.56 \times 10^6$ & 1240 & $9.67 \times 10^5$ \\
S1     & $1.56 \times 10^6$ & 744  & $5.81 \times 10^5$ \\
S2     & $1.56 \times 10^6$ & 495  & $3.86 \times 10^5$ \\
\hline
\end{tabular}
\end{table}
4. Figure S3 shows the MFM images for sample S1 at different applied magnetic field. The labyrinth type of domains appeared at demagnetized state, gradually breaks into skyrmions and then gets saturated at high magnetic field forming a uniform magnetic state.

\begin{figure}[H]
\centering
\includegraphics[width=0.8\linewidth]{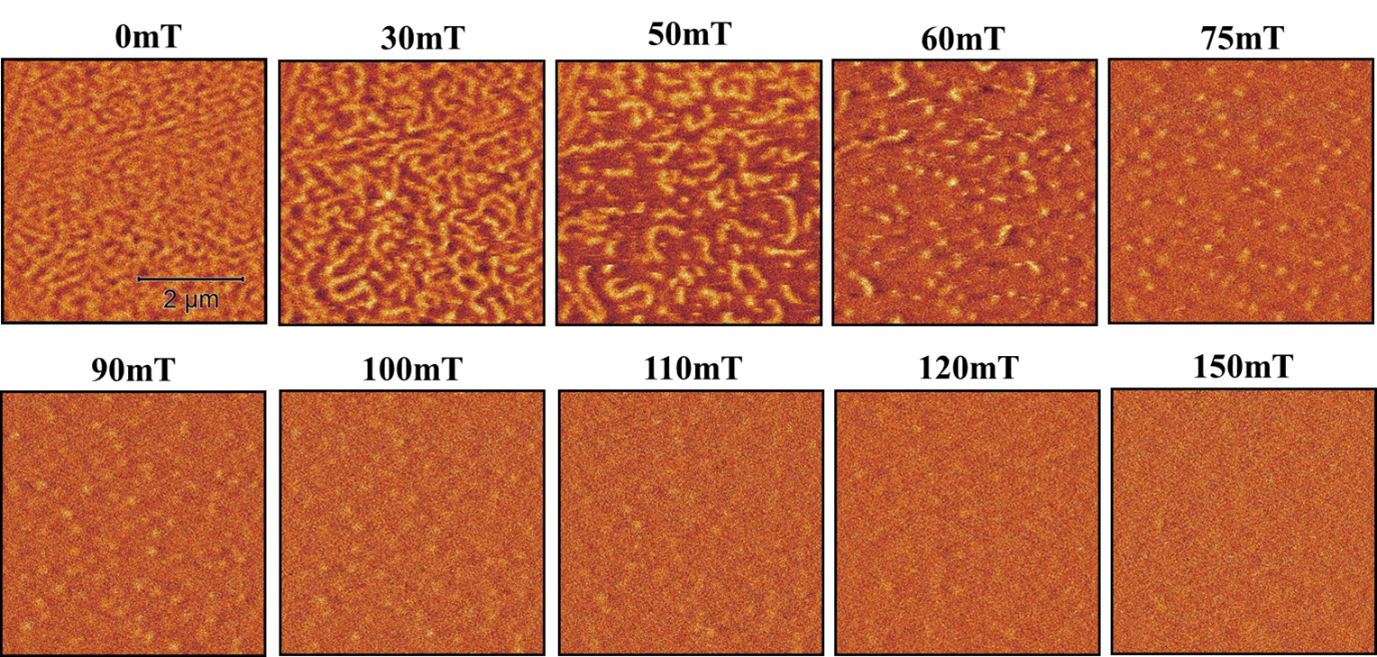}
\caption{Gradual evolution of skyrmionics states from the demagnetized state for S1.}
\end{figure}

5. Figure S4 shows the MFM images for sample S2 at different applied magnetic field. The labyrinth type of domains appeared at demagnetized state, gradually breaks into skyrmions and then gets saturated at high magnetic field forming a uniform magnetic state.

\begin{figure}[H]
\centering
\includegraphics[width=0.8\linewidth]{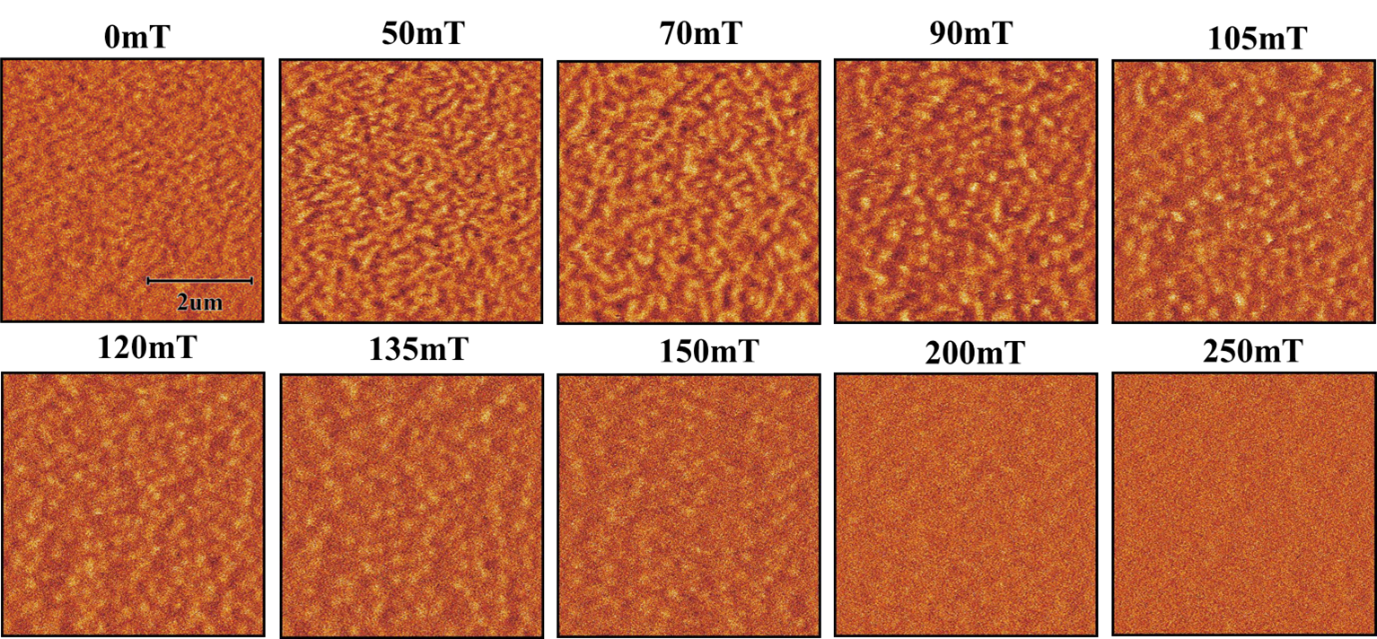}
\caption{Gradual evolution of skyrmionics states from the demagnetized state for S2.}
\end{figure}

6. To measure the Hall resistivity accurately for a rectangular sample, a five-probe method is employed, as shown in Figure S5. While a conventional four-probe setup can measure the Hall voltage using transverse contacts, any misalignment from the ideal perpendicular configuration introduces an additional longitudinal voltage component. To eliminate this, an additional voltage probe is connected on the sample and a potentiometer is used between leads $V_{a^+}$ and $V_{b^+}$. By tuning the potentiometer to ensure $V_{a^+} - V_{b^+} = 0$, the longitudinal contribution is effectively cancelled, allowing precise measurement of the Hall voltage.

\begin{figure}[H]
\centering
\includegraphics[width=0.8\linewidth]{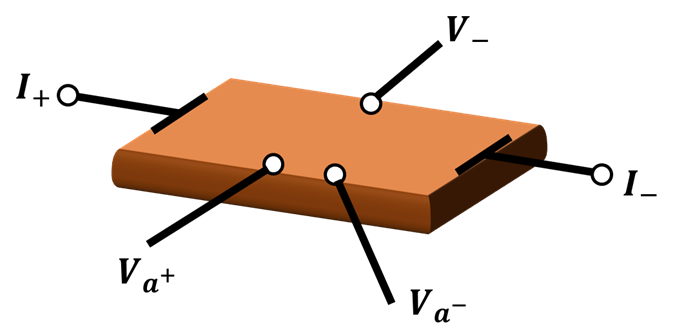}
\caption{Measurement of the Hall resistivity of a rectangular sample by five probe method.}
\end{figure}

\end{document}